\documentstyle[12pt]{article}

\newcommand{\be}{\begin{equation}}
\newcommand{\ee}{\end{equation}}
\newcommand{\ba}{\begin{eqnarray}}
\newcommand{\ea}{\end{eqnarray}}

\begin{document}
\begin{flushleft}
УДК 530.145+004.387
\end{flushleft}

\begin{center}
{\bf
{ Classical and Quantum Problems for Quanputers }}
\footnote{The results of this paper where presented on the
Workshop \\"Quantum physics and communication ", Dubna, May 16-17, 2002.}

{\sl Nugzar Makhaldiani} 

Laboratory of Information Technologies\\[0pt]
Joint Institute for Nuclear Research\\[0pt]
Dubna, Moscow Region, Russia\\[0pt]
e-mail address:~~mnv@jinr.ru
\end{center}


\begin{abstract}
 Quantum computers are considered as a part of the family of the
reversible, lineary-extended,
dynamical systems (Quanputers). For classical problems an operational
reformulation is given. A universal algorithm for the solving of classical
and quantum problems on quanputers is formulated.
\end{abstract}

\vskip 5mm

{\bf 1. Time-reversible classical discrete dynamical systems
(computers) and corresponding (quantum) extraparts
}

The contemporary digital computer and its logical elements can be
considered as a spatial type of discrete dynamical systems:
\begin{eqnarray}  \label{ds}
s_n(k+1)=\phi_n (s(k)),
\end{eqnarray}
where $s_n(k), \ \ 1\leq n\leq N(k),$ is the state vector of the system at
the discrete time step $k$. Note that, with time steps $k,$ can change not
only value, but also the dimension $N(k),$ of the state vector $s(k)$.

{\sl Definition}. We assume that the system (\ref{ds}) is time-reversible if we
can define the reverse dynamical system
\begin{eqnarray}  \label{ids}
s_n(k)=\phi_n^{-1} (s(k+1)).
\end{eqnarray}

In this case the following matrix
\begin{eqnarray}
M_{nm}=\frac{\partial \phi_n(s(k))}{\partial s_m(k)},
\end{eqnarray}
is regular, i.e. has an inverse.
 If the matrix is not regular, this is the case, for example, when
$N(k+1)\neq N(k),$ we have an irreversible dynamical system (usual digital
computers and/or corresponding irreversible gates).

Let us consider an extension of the dynamical system (\ref{ds}) given by the
following action functional:
\begin{eqnarray}\label{A}
A=\sum_{kn}^{}l_n(k)(s_n(k+1)-\phi_n(s(k)))
\end{eqnarray}
and corresponding motion equations
\begin{eqnarray}
&&s_n(k+1)=\phi_n (s(k))=\frac{\partial H}{\partial l_n(k)},\cr
&&l_n(k-1)=l_m(k)\frac{\partial \phi_m(s(k))}{%
\partial s_n(k)} =l_m(k)M_{mn}(s(k))=\frac{\partial H}{\partial s_n(k)},
\end{eqnarray}
where
$H=\sum_{kn}l_n(k)\phi_n(s(k))$,
is discrete Hamiltonian.
 In the regular case, we put this system in an explicit form
\begin{eqnarray}  \label{eds}
&&s_n(k+1)=\phi_n (s(k)),\cr &&l_n(k+1)=l_m(k)M^{-1}_{mn}(s(k+1)).
\end{eqnarray}

From this system it is obvious that, when the initial value $l_n(k_0)$ is
given, the evolution of vector $l(k)$ is defined by evolution of the state
vector $s(k).$ So we have the following

{\sl Theorem.} The regular dynamical system (\ref{ds}) and the extended
system (\ref{eds}) are equivalent.

Note that the corresponding statement from \cite{Birkhoff}
concerning to the continual time dynamical systems is not so
transparent as our statement for discrete time dynamical systems.
In the continual time approximation, the discrete system
(\ref{eds}) reduces to a corresponding continual one, \cite{MV}:
\begin{eqnarray}  \label{sp}
\dot x_n(t)=v_n(x),\ 
\dot p_n(t)=-\frac{\partial v_m}{\partial x_n}p_m.
\end{eqnarray}

Indeed, let us change the dependent variables,
\begin{eqnarray}
&&s_n(k)=x_n(t_k), \ \
l_n(k)=p_n(t_k), \ \ t_k=k\Delta t, \
\ \Delta t<<1.
\end{eqnarray}

Then the action functional (\ref{A}) can be transformed as follows:
\begin{eqnarray}\label{p}
&A&=\sum_{kn}^{}p_n(t_k)(x_n(t_k+\Delta t)-\phi_n(x(t_k)))\cr %
&&\Rightarrow\sum_{kn}^{}\Delta tp_n(t_k)(\dot
x_n(t_k)-v_n(x(t_k)))\cr &&\Rightarrow
\int\limits_{}^{}dtp_n(t)(\dot x_n-v_n(x)),\ \
v_n(t_k)=(\phi_n(x(t_k))-x_n(t_k))/\Delta t,
\end{eqnarray}
and corresponding motion equations are presented by the system
(\ref{sp}). The equation of motion for $l_n(k)$ is linear and has
an important property that a linear superpositions of the
solutions are also solutions.

{\sl Statement.} Any time-reversible dynamical system (e.g. a
time-reversible computer) can be extended by a corresponding
linear dynamical system (quantum-like processor) which is controlled by
the dynamical system and which, due to the superposition and
entanglement properties, has a huge computational power.

Note that \cite{Makh-Shr} for the following (infinite -
dimensional) equation of motion \ba \label{3}iV_t=\Delta
V-\frac{1}{2}V^2, \ea the corresponding linear subsystem is
nothing but the familiar Schr$\ddot o$dinger equation,
\ba\label{4} i\psi_t=-\Delta \psi+V\psi.\ea


The extended system \ba &&iV_t=\triangle V-\frac{1}{2}V^2,\cr
&&i\psi_t=-\Delta \psi+V\psi,
 \ea
 is mathematically and physically complete, as potential is not an
 external function, but is part of the Hamiltonian dynamical model.

{\bf 2. Solution of the Quantum problems with the Methods of the
Quantum Computing}

The standard classical computing - the technology of storing and
transforming information, is based on the classical physical
theory and can be universally divided (factored) on the hardware
(body) - memory and processor of computers, and software (soul) -
algorithms and programs of users, parts.
Under the Quantum computing (Quanputing, QC), we mean an extension
of the classical computing by hardware which is described by the
(corresponding extention of the) quantum theory - quantum
registers and quantum logical units and corresponding software.
In Quanputing the factorization on hardware and software is only
an approximate concept, $
<SH>=<S><H>+O(\hbar).$

A quantum system can be described by corresponding Schr$\ddot o$dinger
equation \cite{Berezin}
\begin{eqnarray}  \label{SE}
i\hbar \frac{d}{dt}|\psi>=\hat H|\psi>,
\end{eqnarray}
where $|\psi>=|\psi(t)>$ is the state vector from the state
Hilbert space and $\hat H=H(\hat p,\hat x)$ is an
operator-Hamiltonian. In the case of one nonrelativistic particle
the operator is
$\hat H=\hat T+\hat V=\frac{\hat p^2}{2m}+V(\hat x).$
The fundamental bracket is $[\hat x,\hat p]=\hat x\hat p-\hat
p\hat x=i\hbar.$  The configuration space form of Eq. (\ref{SE}) is
\begin{eqnarray}  \label{xSE}
i\hbar \frac{\partial\psi(x,t)}{\partial t}=\hat H\psi(x,t),
\end{eqnarray}
where $\psi(x,t)=<x|\psi(t)>$ and Hamiltonian is
$\hat H=-\frac{\hbar^2}{2m}\frac{d^2}{dx^2}+V(x).$

The formal solution of Eq. (\ref{SE}) is
\begin{eqnarray}  \label{FS}
|\psi(t)>=U(t)|\psi_0>,\ \
U(t)=exp\ (-\frac{i}{\hbar}t\hat H).
\end{eqnarray}
The main steps maid in QC are the following:
\begin{eqnarray}  \label{UFa}
U(t)=(U^{1/N})^N=(U_TU_V)^N+O(1/N),
\end{eqnarray}
where
\begin{eqnarray}  \label{Ufac}
U^{1/N}&=&exp\ (-\theta \hat H)=exp\ (-\theta \hat T) exp\
(-\theta \hat V)+O(1/N^2)\cr &=&U_TU_V+O(1/N^2),\ \
\theta=\frac{i}{\hbar}\tau,\ \ \tau=\frac{t}{N}.
\end{eqnarray}

Then, for a corresponding matrix element we have (see Appendix)
\begin{eqnarray}
<x_{n+1}|U^{1/N}|x_n>\sim exp\
(\theta(\frac{m}{2}(\frac{x_{n+1}-x_n}{\tau})^2 -V(x_n)))+O(1/N^2)
\end{eqnarray}
and
\begin{eqnarray}
<x_{out}|U(t)|x_{in}>&\sim &\int\limits_{}^{}dx_1dx_2...dx_N exp\ (\frac{i}{%
\hbar}\tau \sum_{n=0}^{N}(\frac{m}{2}(\frac{x_{n+1}-x_n}{\tau})^2-V(x_n)))%
\cr &+&O(1/N),\ \ |x_0>=|x_{in}>,\ \ <x_{N+1}|=<x_{out}|.
\end{eqnarray}

This finite dimensional integral representation of the matrix
element is in the ground of the functional (continual) integral
formulation of the quantum theory \cite{Feynman}. Then we
discretize the wave function in (\ref{xSE}), (\ref{FS}) and impose
the periodic boundary conditions \cite{Zalka}
$a_n(t)=\psi(x_n,t), \ \ a_{n+N}=a_n,\ \ x_n=n\Delta x.$

We store these amplitudes in a k-bit quantum register,
\begin{eqnarray}
|\psi(t)>=\sum_{n=1}^{N}a_n|n>,\ \ N=2^k,
\end{eqnarray}
where $|n>$ is the basis state corresponding to the binary representation of
the number n.
The second factor in (\ref{Ufac}) in a coordinate representation corresponds
to a diagonal unitary transformation of the quantum computer state $%
\psi(x,t).$ After Fourier transformation $\psi(x,t)$ into momentum-space
representation $\psi(p,t)$ the first factor in (\ref{Ufac}) can be applied
in the same way.
Diagonal unitary transformations of the type
$|n>\rightarrow e^{iF(n)}|n>,$
where $F(n)$ is a function of n, can be done \cite{Zalka} with the
following succession of steps
\begin{eqnarray}
|n>&\rightarrow &|n,0>\rightarrow |n,F(n)>\rightarrow e^{iF(n)}|n,F(n)>
\rightarrow e^{iF(n)}|n,0>\cr &\rightarrow &e^{iF(n)}|n>.
\end{eqnarray}

For n particles in d dimensions we need $nd$ quantum registers. If
the potential $V(\hat x_1,\hat x_2,...)$ couples different degrees
of freedom, we need the diagonal unitary transformations acting on
several registers,
\begin{eqnarray}
|n_1,n_2,...>\rightarrow e^{iF(n_1,n_2,...)}|n_1,n_2,...>.
\end{eqnarray}

The disctetized bosonic quantum fields are naturally defined on
the space-time lattice, scalar (particles) fields - on lattice
sites, vector particles - on lattice links, tensor particles - on
lattice plackets, and so on (see e.g. \cite{CFT}).

{\bf 3. Solution of the Classical Problems with the Methods of the
Quantum Computing}

If we can solve the quantum problems with the quantum computers
\cite{QCR}, we can probably solve the classical problems, too.
With the paper \cite{Makh-137} we started the investigation of
computational hard problems of classical physics, such as the
turbulent phase of hydrodynamics, with the methods of Quantum
Computing (QC).


We usually formulate the classical problems as a Hamiltonian system of motion
equations \cite{fad}
\ba \label{HE}
\dot{q}_{n}=\frac{\partial H}{\partial p_n},\ \
\dot p_n=-\frac{\partial H}{\partial q_n},\ \ \ 1\leq n\leq N,
\ea
where $q_{n}$ are coordinates for the configuration space of the system and
$p_{n}$ are corresponding momentum, $H=H(q,p)$ is a Hamiltonian function(al).
The system (\ref{HE}) belongs to a more general class of the dynamical
systems defined by the following system of motion equations
\ba \label{DS}
\dot{x}_{n}=v_{n}(x),\ \ 1\leq n\leq N,
\ea
when N is an even integer and
\[
v_{n}(x)=\varepsilon _{nm}\frac{\partial H(x)}{\partial x_{m}}=\{x_{n},H\},
\]
with the fundamental canonical bracket
$\{x_{n},x_{m}\}=\varepsilon _{nm}.$

 Note that any dynamical
system (\ref{DS}) can be extended to the following Hamiltonian
form, (see, e.g. \cite{MV} and section 1 of this paper.) \ba
&&\dot{x}_{n}=v_{n}(x)=\{x_{n},H\},\cr
&&\dot{p}_{n}=-\frac{\partial v_{m}}{\partial
x_{n}}p_{m}=\{p_{n},H\}, \ea with Hamiltonian
$H(x,p)=\sum_{n}v_{n}(x)p_{n}$ and the fundamental canonical
bracket $ \{x_{n},p_{m}\}=\delta _{nm}.$

For any observable $\psi(x)$ of the Hamiltonian dynamical system,
we have the following motion equation:
\begin{eqnarray}  \label{HSE}
\dot \psi(x)=\{\psi,H_1\}=\varepsilon_{nm}\frac{\partial \psi}{\partial x_n%
} \frac{\partial H_1}{\partial x_m}=-\hat L\psi
=-\frac{i}{\hbar }\hat H_2\psi,
\end{eqnarray}
where
\begin{eqnarray}
\hat L=v_n(x)\frac{\partial }{\partial x_n}=\frac{i}{\hbar}v_n(x)\hat p_n %
=\frac{i}{\hbar}\hat H_2,
\end{eqnarray}
$H_1$ is the classical Hamiltonian function and $\hat H_2$ is the quantum
Hamiltonian operator. Now we can consider Eq.(\ref{HSE}) as a
classical Hamiltonian one with Hamiltonian $H_1$ or as a quantum Schr$%
\ddot o$dinger equation with the Hamiltonian operator $\hat H_2.$
For the second form we can apply the QC methods developed in the
previous section.

Note that the operator $\hat H_2$ is a self-adjoint operator, $\hat H_2^+=
\hat H_2,$ when $divv=0$ and we can put it in the second-order form with
respect to the momentum operator,
\begin{eqnarray}
\hat H_2=v_k\hat p_k=\frac{1}{2}(\hat p_kv_k+v_k\hat p_k)
=\frac{i}{%
\hbar}[\hat p_k^2,u_k],
\end{eqnarray}
where
\begin{eqnarray}
u_k=\frac{1}{2}\int\limits_{}^{x_k}dx_kv_k, \ \ 1\leq k\leq d,\ \ d\geq 2.
\end{eqnarray}


The formal solution (\ref{FS}) is
\[
|\psi (t)>=U(t)|\psi _{0}>=exp\ (-\frac{i}{\hbar
}t\hat{H}_{2})|\psi _{0}>=exp\ (-\frac{t}{\hbar
^{2}}[\hat{u},\hat{p}^{2}])|\psi _{0}>
\]
and for the factorized form (\ref{UFa}) we have
\ba\label{cUfac}
U^{1/N}=exp\ (-\tau ^{2}[\hat{u},\hat{p}^{2}])=\prod_{k=1}^{d}e^{i\tau \hat{u}%
_{k}}e^{i\tau \hat{p}_{k}^{2}}e^{-i\tau \hat{u}_{k}}e^{-i\tau \hat{p}%
_{k}^{2}}+O(\tau ^{3}),
\ea
where
\[
\tau =\pm (\frac{t}{\hbar ^{2}N})^{1/2}.
\]

Now we are ready to apply the formalism of the QC considered in Sec.3 to the
classical problems.
In the extended version of this paper, we describe some
computational problems from different parts of physics.
The Hamiltonian dynamics of heavy-particle accelerators; the
standard model of condensed state physics (Hubbard model); the 123
standard model of high-energy (elementary particles) physics;
unified string and M-theory models.

{\bf 4. Conclusions and perspectives}

In this paper we constructed an algorithm for solving difficult
computational(hard) problems of classical and quantum physics on
quanputers.
Note that before we get a real quanputer, we can simulate it
on classical computers \cite{DR}.
It is interesting to investigate the functional integral
formulation of the classical theory based on a discrete
representation (\ref{cUfac}). Then, Dirac's equation for electron
\begin{eqnarray}
i\hbar \partial_t\psi (t,x)=(\alpha_n\hat p_n+\beta m)\psi(t,x)
=(%
\frac{i}{\hbar }[\hat p_n^2,u_n]+\beta m)\psi,
\end{eqnarray}
(where
$u_n=\frac{1}{2}\int\limits_{}^{}dx_n\alpha_n=\alpha_nx_n/2,$)
for $m=0,$ has the form similar to the classical equation
(\ref{HSE}), (with $\alpha_n$ as velocity matrices) and the
Planck's constant $\hbar$ can be cancelled. This is an inverse to
the consideration made when we try to find quantum (photon) nature
of the (classical) Maxwell's equations and put them in the
"Dirac's form", (see e.g. \cite{Bjorken}). It is possible to
formulate similar equations for high spin massless fields.
 It is interesting to find a corresponding classical
interpretation of these equations and make observable
predictions.
The effects of two of them (the
photon's and graviton's) are obvious in everyday life.
What about the effects of other massless particles?







\bigskip
{\bf Appendix }

{\bf 1. }In the main text of this paper we used the following
relation:
\begin{eqnarray}
e^{\varepsilon A}e^{\varepsilon B}e^{-\varepsilon A}e^{-\varepsilon B}=
e^{\varepsilon^2[A,B]}+O(\varepsilon^3).
\end{eqnarray}

The relation
\begin{eqnarray}
(1+\varepsilon A)(1+\varepsilon B)(1+\varepsilon
A)^{-1}(1+\varepsilon B)^{-1}
=(1+\varepsilon^2[A,B])+O(\varepsilon^3),
\end{eqnarray}
 may also be useful.

{\bf 2.} Coordinate and momentum state vectors are correspondingly
$|x>$ and $|p>,\ \hat x|x>=x|x>,\ \ \hat p|p>=p|p>,$
\begin{eqnarray}
 &&<p|x>=\psi_x(p)= \frac{1}{\sqrt{%
2\pi\hbar}}\ exp\ (-\frac{i}{\hbar}px),\ \hat x\psi_x(p)=i\hbar\frac{\partial%
}{\partial p}\psi_x(p)=x\psi_x(p),\cr &&<x|y>=\int\limits_{}^{}dp<x|p><p|y>=%
\int\limits_{}^{}\frac{dp}{2\pi\hbar}\ exp\
(\frac{i}{\hbar}p(x-y))\cr && =\delta (x-y).
\end{eqnarray}
 Now we calculate the following matrix element
\begin{eqnarray}
<x_{n+1}|\ exp~(-a\hat
p^2)|x_n>&=&\int\limits_{}^{}d^Dp<x_{n+1}|p><p|x_n>
exp\ (-ap^2)\cr =\int\limits_{}^{}\frac{d^Dp}{(2\pi \hbar)^D}\ exp\ (i\frac{%
p(x_{n+1}-x_n)}{\hbar}-ap^2)
&=&\frac{A^D}{(2\pi\hbar)^D}\ exp\ (-\frac{%
(x_{n+1}-x_n)^2}{4a\hbar^2}),
\end{eqnarray}
where in the case of the quantum mechanics of the particle, (\ref{Ufac})
\begin{eqnarray}
a=i\frac{t}{2m\hbar N}
\end{eqnarray}
and in the case of classical mechanics, (\ref{cUfac})
\begin{eqnarray}
a=i(\frac{t}{\hbar^2N})^{1/2},
\end{eqnarray}
\begin{eqnarray}
A=\int\limits_{}^{}dp\ exp\ (-ap^2)=\sqrt{\frac{\pi}{a}}.
\end{eqnarray}





\begin{thebibliography}{99}
\bibitem{Birkhoff} {\sl Birkhoff G.D.}  Dynamical Systems, Amer. Math.
Soc.,\\
Providance, R.I., 1927.
\bibitem{MV} {\sl Makhaldiani N., Voskresenskaya O.} JINR Commun. E2-97-418.\\
 Dubna, 1997.
\bibitem{Makh-Shr} {\sl Makhaldiani N.} JINR Commun. E2-2000-179. Dubna, 2000.
\bibitem{Berezin}  {\sl Berezin F., Shubin M.}  Schr$\ddot{o}$dinger equation, Kluwer, Dordrecht, 1991.
\bibitem{Feynman}  {\sl Feynman R.P., Hibbs A.R.}  Quantum Mechanics and
Path Integrals. McGraw-Hill, New York, 1965.
\bibitem{Zalka} {\sl  Zalka C.}  Proc. Roy. Soc. Lond. 1998. A454. P. 313
\bibitem{CFT}  {\sl Makhaldiani N.M.}  Computational Quantum Field Theory,\\ JINR Commun. P2-86-849. Dubna, 1986.
\bibitem{QCR}  {Ecert A., Jozsa R.} Rev. Mod. Phys. 1996. V. 68, P. 115
(1996).
\bibitem{Makh-137}{\sl Makhaldiani N.} JINR Commun. E2-2001-137. Dubna, 2001.
\bibitem{fad} {\sl Faddeev L.D., Takhtajan L.A.}  Hamiltonian methods in
the theory of solitons, Springer, Berlin, 1987.
\bibitem{DR} {\sl Hans De Raedt et al.}  Comp. Phys. Com. 2001. V. 132. P. 1
\bibitem{Bjorken} {\sl  Bjorken J.D., Drell S. D.}  Relativistic
Quantum Mechanics, McGraw-Hill, New York, 1965.
\end{thebibliography}
\end{document}